\documentclass[12pt,oneside]{amsbook}
\usepackage[margin=1in]{geometry}
\usepackage{mathptmx}

\newtheorem*{definition*}{Definition}
\newtheorem*{theorem*}{Theorem}
\newtheorem{theorem}{Theorem}[section]
\newtheorem{conjecture}[theorem]{Conjecture}
\newtheorem{proposition}[theorem]{Proposition}
\newtheorem{corollary}[theorem]{Corollary}
\newtheorem{lemma}[theorem]{Lemma}

\newcommand{\order}[1]{\lvert#1\rvert}
\newcommand{\size}[1]{\lVert#1\rVert}
\newcommand{\ceiling}[1]{\lceil#1\rceil}
\newcommand{\floor}[1]{\lfloor#1\rfloor}
\newcommand{\braket}[1]{\langle#1\rangle}
\newcommand{\slog}{\operatorname{slog}}

\begin{document}

\frontmatter

\begin{center}
No Quantum Brooks' Theorem\\
By\\
STEVEN LU\\
B.A. (University of California, Berkeley) 2009\\
DISSERTATION\\
Submitted in partial satisfaction of the requirements for the degree of\\
DOCTOR OF PHILOSOPHY\\
in\\
MATHEMATICS\\
in the\\
OFFICE OF GRADUATE STUDIES\\
of the\\
UNIVERSITY OF CALIFORNIA\\
DAVIS\\
Approved:\\
\rule{3in}{1pt}\\
Greg Kuperberg, Chair\\
\rule{3in}{1pt}\\
Eric Babson\\
\rule{3in}{1pt}\\
Bruno Nachtergaele\\
Committee in Charge\\
2014\\
\end{center}

\tableofcontents

\chapter*{Abstract}

First, I introduce quantum graph theory. I also discuss a known lower bound on the independence numbers and derive from it an upper bound on the chromatic numbers of quantum graphs. Then, I construct a family of quantum graphs that can be described as tropical, cyclical, and commutative. I also define a step logarithm function and express with it the bounds on quantum graph invariants in closed form. Finally, I obtain an upper bound on the independence numbers and a lower bound on the chromatic numbers of the constructed quantum graphs that are similar in form to the existing bounds. I also show that the constructed family contains graphs of any valence with arbitrarily high chromatic numbers and conclude by it that quantum graph colorings are quite different from classical graph colorings.

\mainmatter

\chapter{Introduction}

There are two halves of the theory behind quantum computers. One of those halves is quantum computation, the quantum equivalent of the work of Turing and others. Included in quantum computation is the study of quantum algorithms, such as Grover's and Shor's algorithms, as well as the study of quantum gates and circuits. The other half is quantum information, the quantum equivalent of the work of Shannon and others. Included in quantum information is the quantum codes, such as the Shor and Steane codes, as well as the study of quantum operations and noise. It is within quantum information that this thesis firmly lies. The tools used in this thesis can be broadly categorized as operator theory, coding theory, or graph theory, corresponding roughly to the fields of physics, computer science, and mathematics which all contribute to quantum information theory. As such, there are three different stories to tell before I can present my results.

There are many equivalent ways to present the principles of quantum mechanics. One such way is as a non-commutative probability theory which is consistent with the Copenhagen interpretation. In quantum probability, a probabilistic system has an algebra of random variables that forms a von Neumann algebra; these algebras were first described by von Neumann, who, with Murray, showed that they are generalizations of classical measure spaces \cite{MN}. Similarly, Weaver shows that von Neumann algebras can be used to define a quantum generalization of classical relations \cite{NW}. The interpretation of a von Neumann algebra in quantum mechanics is as an algebra of observables corresponding to a quantum system. In Heisenberg's matrix mechanics, observables are bounded linear operators on the Hilbert space of states. Its relation to the von Neumann algebra approach is revealed by a theorem of Gelfand and Naimark which implies that every von Neumann algebra is isomorphic to a subalgebra of bounded linear operators on a Hilbert space \cite{GN}. This is unsurprising when considering that von Neumann originally defined them as subalgebras, but Sakai showed that von Neumann algebras can be characterized algebraically and independent of representation \cite{SS}. While the fully classical systems are easily described as those with commutative von Neumann algebras, fully quantum systems are more complicated in nature. The most non-commutative von Neumann algebras can be classified as one of three types, but only one of the types includes finite dimensional algebras. The Artin-Wedderburn theorem implies that finite dimensional von Neumann algebras are least commutative when they are full matrix algebras \cite{JW}. Then, the proof of Choi's theorem gives a very manageable form for a fully quantum channel \cite{MC}.

The study of channels is called information theory. A fundamental result of information theory is that all classical information is quantifiable; this quantity is called entropy. The measurement of entropy and the entropy capacity of a channel were defined by Shannon, who, in the same article, described a general model of a noisy classical channel \cite{CS}. While Shannon gave an upper bound on the capacity of a channel, a reliable method of communicating with a noisy channel at all was not known until Hamming's discovery of error correcting codes \cite{RH}. Additionally, Hamming proved an upper bound on the efficiency of such codes and found a family of codes for which the bound is sharp. Many channels, however, do not allow for perfect codes, so finding upper and lower bounds on code size are important open problems. Similar to a classical error correcting code is a quantum error correcting code, such as the Shor code, which can effectively transmit quantum information over a noisy quantum channel \cite{PS}. Knill and Laflamme give a better characterization of quantum codes that is independent of a recovery scheme and therefore defined solely by its elements \cite{KL}. They adapt much of classical error correction, such as the Hamming bound, to use for quantum error correction. Some classical codes have been adapted to quantum codes as well. An example of this is the large class of quantum codes called additive codes, discovered by Calderbank, Rains, Shor, and Sloane, which are the quantum analog of classical linear codes \cite{CRSS}. Classical and quantum error correction are generalized by Kuperberg and Weaver to the von Neumann algebra setting by generalizing classical metric spaces, and this leads to a definition of quantum graphs \cite{KW}.

The bounding of invariants is a significant part of graph theory. Three such graph invariants are independence number, Shannon capacity, and chromatic number. A pair of independence number bounds, called the packing and covering bounds, are functions of the graph's order and valence; the sharpness of the covering bound is proved by Tur\'{a}n's theorem \cite{PT}. The Lov\'{a}sz number of a graph is a computable bound on the Shannon capacity that also bounds other invariants \cite{LL}. The chromatic number of the complement graph, for instance, is an invariant that is nontrivially bounded by the Lov\'{a}sz number. The chromatic number has varied applications and a very rich theory. One surprising result is a theorem of Erd\H{o}s which states that, while trees have chromatic number two, graphs that are only locally trees (i.e., high girth) can have arbitrarily high chromatic number \cite{PE}. One unsurprising upper bound on chromatic number is one more than the maximum valence, which can be achieved by a greedy coloring, but the exact cases of its sharpness is proven by Brooks \cite{RB}. In the same sense as other quantum objects, a quantum graphs is defined by a non-commutative graph theory which can be mutually generalized with classical graph theory. However, quantum graph theory has not seen much study. One of two previous results on quantum graphs is the bound on independence number based on order and valence by Knill, Laflamme, and Viola, which was not presented as quantum graph theory \cite{KLV}. A quantum Lov\'{a}sz number, bounding the quantum Shannon capacity, has been studied as quantum graph theory by Duan, Severini, and Winter, and the chromatic number of quantum graphs has never been defined \cite{DSW}.

The intersection of these three topics is quantum graph theory. It is defined by operator theory, motivated by coding theory, and guided by graph theory. In this spirit, quantum graph theory is equivalent to the most non-commutative finite dimensional case of the von Neumann algebra generalization of graph theory provided by the generalization of metric spaces and therefore of graph metrics. The consequence of this definition is that, as in the classical case, every quantum graph defines a quantum channel whose quantum codes are exactly the independent sets of the defining quantum graph. As with other quantum analogs, a lot of classical graph theory translates closely, including very similar looking bounds on independence number and a very similar looking Lov\'{a}sz number. However, a quantum graph coloring, as I define it, can not be bounded by a function on valence only, in contrast with the Brooks bound. This is Theorem \ref{chromatic}. Another result is Theorem \ref{independence}, an upper bound on the size of a family of quantum graphs. The method of proof can be briefly described as exploiting the large difference between the median and the mean of a tropically distributed set of eigenvalues--by defining a family of tropical quantum graphs and proving lemmas restricting code size by both medians and means of eigenvalues, the family can be shown to contain graphs of any valence and arbitrarily high chromatic number.

\chapter{Classical Channels}

A good review of classical error correction makes quantum error correction seem very natural. Classical channels are presented here first as maps, then as relations, and finally as graphs. The definitions of states, error correcting codes, and error detecting codes are introduced along the way, and examples of each are given. Three important properties of graphs--independence number, Shannon capacity, and chromatic number--are then discussed, with upper and lower bounds given for each. These concepts are generalized in the next chapter.

\section{As Maps}

A classical channel models the transfer of classical information \cite{CS}. In one use of a classical channel, a message is sent through the channel, and a message is then received. If an error occurs, the message received is not the message sent. There is a finite set of messages that can be transmitted through a channel, and classical information is always modeled as an element of a set rather than as one of infinitely many possibilities. This apparent conflict with the typical perception of information is resolved by sets of alphabets and product channels, which will be discussed later. It should also be noted that the messages themselves are irrelevant, and the only quantity determining the size of a channel is the number of distinct messages that can be transmitted. For a given channel, let $p(a,b)$ denote the probability of sending a message $a$ and receiving a message $b$; this is called the transition probability. Since probability must always be conserved, $$\sum_b p(a,b)=1$$ for any message $a$. Organizing these transition probabilities into a matrix $M$ with rows and columns indexed by messages such that $$M_{a,b}=p(a,b)$$ gives a stochastic matrix, and this is the standard definition of a classical channel. The dimensions of the matrix are both equal to the size of the message set.

\begin{definition*}A classical channel is a stochastic matrix, which is a nonnegative square matrix with each column summing to $1$.\end{definition*}

Let $X$ be the set of messages of a channel, which are also called pure states. If the messages are represented as elementary basis vectors in $\mathbb{R}^X$, then the result of sending a message through the channel is given by the left multiplication of the channel's stochastic matrix. The resulting vector will have entries that are nonnegative probabilities summing to $1$ and not necessarily basis vectors. These vectors are called mixed states, and they represent statistical ensembles of messages. The set of all states, both pure and mixed, is called the state space; it is a simplex of dimension one less than the number of messages. Notice that the state space is also a convex set whose extremal points are exactly the pure states. If $\mathbb{R}^X$ is equipped with the $1$-norm, then the state space is simply the intersection of the unit sphere with the nonnegative orthant. This description of the state space will be useful later in generalizing the concept of states. The characterization of linear operators on the state space is simple--they correspond exactly to the left multiplications by stochastic matrices. The nonnegativity of the matrix preserves the nonnegativity of the states, and the column sum condition preserves the $1$-norm of the states. Then a classical channel can also be viewed as a description of how classical states can evolve and can alternatively be defined as any linear operator on the state space. The restriction of linearity on maps between states is called classical superposition, and it follows from classical probability.

\begin{definition*}A classical state is a vector on which a channel acts; its entries must be nonnegative and must sum to $1$.\end{definition*}

An example of a classical channel is the memory of a classical computer. In a classical computer, all information is converted to and stored as a string of bits, which are classical channels that have message sets of size $2$; these messages are commonly called $0$ and $1$. A string of $n$ bits, therefore, has a message set of size $2^n$. Such a set is called a Hamming space. Some time after a string is stored, it is read from the memory, so the sender and receiver are the same computer in this case. One type of error that could affect this channel is a bit flip, which causes any single bit in the string to change value. Suppose a channel stores strings of length $N$, and every single bit flip independently has the same probability $P$ of occurring. Any two strings of length $N$ can be connected by some sequence of bit flips, so $p(a,b)$ is nonzero for any two strings $a$ and $b$ of the same length if $P$ is nonzero. However, the transition probability will be much higher between strings with many matching bits. Then the $2^N$ by $2^N$ stochastic matrix $M$ describing the channel is defined by $$M_{a,b}=P^{k(a,b)}(1-P)^{N-k(a,b)},$$ where $k(a,b)$ is the number of bits in which $a$ and $b$ differ. The column sums, by the binomial theorem, are all $1$. Note that, for small $P$, the probability of more than one bit flip is far less than the probability of exactly one bit flip. This provides an obvious choice of threshold for the simplification made in the next section.

\section{As Relations}

For a channel to be useful, there must be a way to detect or correct errors \cite{RH}. It is often impossible to eliminate every error, such as when $p(a,b)$ is nonzero for every $a$ and $b$. Then, any message received can correspond to any message sent, so no information can be transmitted if no errors are allowed. In some cases, there is simply a distinguished set of errors that must be corrected, and all other errors are ignorable. In all cases, a threshold probability of errors to correct can be considered. Any error with probability above the threshold is corrected, and any error with probability below the threshold is seen as too unlikely to need correction. This simplifies a classical channel to a set of messages and a set of probable errors between them, which defines a relation. One way to do this is to round the entries in a channel's stochastic matrix to either $0$ or $1$, depending on which side of the threshold probability the entry lies. The resulting $0$-$1$ matrix $R$ acts on the state space in a natural way: two pure states $a$ and $b$ are unrelated if and only if $$a^TRb=0.$$ A matrix relation can also be extended to mixed states, but it amounts to no more than determining whether any of the messages in the support of each mixed state are related because the relation condition extends linearly. This is the definition of a relation that will be generalized in the next chapter, but a classical relation is most often defined as a set. Given a channel and its stochastic matrix $M$, define the error relation as $$\{(a,b):M_{a,b}\geq P\}$$ for some chosen threshold probability $P$.

\begin{definition*}A classical relation $R$ on a set $X$ is a subset of $X\times X$; say that $a$ is related to $b$ if $(a,b)\in R$.\end{definition*}

An error correcting code of a channel is a subset of messages. Error correction with an error correcting code is as follows. Only messages in the code are sent through the channel. From a received message $b$, the original message can be recovered if there is an unique element $a$ of the code such that $(a,b)$ lies in the relation, since the relation contains all probable transitions. If the code is chosen such that the sets $$R(a):=\{b:(a,b)\in R\}$$ are disjoint for every $a$ in the code, then there will be a unique sent message corresponding to each possible received message, and error correction is complete. By using only a subset of the messages available, however, the size of the channel has effectively been reduced. The amount of size that must be sacrificed to correct errors depends on how many errors need to be corrected. An error correcting code encoding $n$ bits into an $m$ bit channel is called a $(m,n)$ code. Given any choice of a code, there is a set of errors that it corrects, of which the original set of errors should be a subset for error correction to succeed. Then our simplification earlier of a channel to an error relation is justified. Rather than specifying the exact probabilities of each error as a stochastic matrix, the errors to be corrected can be specified as a relation because the error correcting scheme does not involve probabilities. Any choice of error correcting code must result from a choice of errors to correct, so making such a choice by threshold probability or otherwise is necessary.

\begin{definition*}A classical error correcting code $C$ of a channel with relation $R$ is a subset of messages such that $R(a)\cap R(b)=\emptyset$ for any $a,b\in C$.\end{definition*}

One example of an error correcting code is the repetition code, and it can correct any single bit flip. If the threshold probability chosen is equal to the probability of exactly one bit flip, then the bit flip channel discussed in the last section becomes a relation $R$ defined by $(a,b)\in R$ if and only if $a$ and $b$ differ at most by one bit. Using a repetition code, a $3N$ bit channel can send $N$ bit messages by repeating the message twice, and single bit flips can be corrected because the $N$ bit blocks can be checked against each other for errors. If at most a single bit flip has occurred, then at least two of the three blocks will agree with each other, and they will contain the correct message. For instance, the message $01$ can be transmitted as $010101$. If a bit flip error affects the third bit, then the message is received as $011101$. The three blocks--$01$, $11$, and $01$-- will reveal the transmitted message by a majority vote. Therefore, the repetition code is a $(3N,N)$ code. Note that a channel can also communicate half of its bits by repeating once, but this code would only be able to detect the presence of an error if the two halves do not match without being able to correct the error because there is no majority. This is called an error detecting code, and it is defined in the next section.

\section{As Graphs}

A relation is a very general model of a channel, and a more manageable model is a graph \cite{RB}. This requires two more simplifications to be made. The first is reflexivity. A relation $R$ is reflexive if $(a,a)\in R$ for every $a$ in the message set. This is merely assuming that the channel has enough fidelity for the probability of not having an error to be at least the threshold probability. For many of the channels worth considering, this is the case. The second assumption is symmetry. A relation $R$ is symmetric if $(a,b)\in R$ implies $(b,a)\in R$ for any messages $a,b$. In the example of bit flip errors and in many other cases, it is already true. Otherwise, symmetry can be achieved by adding in the necessary relations. A reflexive, symmetric relation is then equivalent to a graph with vertices corresponding to messages and edges corresponding to errors in the relation. As with classical relations, a classical graph can be decribed by a matrix that is a rounding of the channel's stochastic matrix. Reflexivity of the relation implies that the diagonal entries are all $1$, and symmetry of the relation implies that the transpose of the matrix should be equal to itself. This is also the definition that will be generalized, but a classical graph is usually defined in a different way. Define the confusability graph of a channel with relation $R$ as $$\{\{a,b\}:(a,b)\in R,a\neq b\},$$ which is equal to $$\{\{a,b\}:M_{a,b}\geq P,a\neq b\}$$ for the channel's stochastic matrix $M$ and threshold probability $P$. Note that the self loops are excluded in this definition, which allows a rigorous definition of an error. The self loops are not errors, and every other pair is called an error.

\begin{definition*}A classical graph $G$ on a set $X$ is a set of unordered pairs of elements in $X$ that includes all singletons; the elements of $G$ are called edges, and the elements of $X$ are called vertices.\end{definition*}

An error detecting code on a channel is a subset of messages, much like an error correcting code. In fact, any error correcting code is an error detecting code, since any code that can correct errors can trivially also detect errors by comparing the corrected message and the received message. For an error detecting code, however, the original message is not recovered if an error occurs. As in error correction, only messages in the error detecting code are transmitted. If the received message is not in the code, then an error has occurred. If the code is chosen such that no two different elements share an edge, then any message received that is in the code has not incurred an error. This explains the earlier assumption of symmetry because it only matters that no code elements share an error and not in which direction the error occurs; detection in one direction implies detection in the other direction as well. Error correcting codes, as described in the previous section, also have a simple representation on the confusability graph: they are the distance three sets. That is, they are subsets such that each element is at least three edges apart from any other element. If this holds, then the neighborhoods of each element of the code are disjoint, and the error correcting procedure works. Error detecting codes are more general than error correcting codes in the following sense. Finding an error correcting code of a graph is equivalent to finding an error detecting code of the graph with an edge added between the endpoints of any length two path of the original graph; such a graph is called the square of the original graph. Thus, an error detecting code simply detects twice as many errors as an error correcting code corrects in some sense, and they often do not need to be considered separately. Therefore, I will use the word ``code'' to refer to error detecting codes.

\begin{definition*}A classical error detecting code $C$ on a channel with graph $G$ is a subset of messages such that $\{a,b\}\not\in G$ for any distinct $a,b\in C$.\end{definition*}

A more advanced example of a code is the Hamming code. The Hamming code encodes $2^k-k-1$ bits into $2^k-1$ bits and corrects up to one single bit flip or detects up to two bit flips. The $(7,4)$ Hamming code, as an example, works by using the first, second, and fourth bit as parity bits, which are bits not encoding the message but included to protect against errors. Given a four bit message, a codeword is generated by setting its third, fifth, sixth, and seventh bit equal to the message. The first bit is then set to the XOR of the third, fifth, and seventh bits. The second bit is set to the XOR of the third, sixth, and seventh bits. The fourth bit is set to the XOR of the fifth, sixth, and seventh bits. For instance, the message $0101$ is encoded as $01\underline{0}0\underline{101}$, where the message bits are underlined. A received message is checked in the first, second, and fourth bits in the same way it was sent. If a bit flip error occurs on a message bit, then at least two of the parity checks will be incorrect; which checks fail will give the location of the error. If a bit flip error occurs on a parity bit, then only one parity check will fail, and that will be the location of the error. In fact, from the choice of parity bits, the three parity checks concatenated in reverse order give the binary representation of the numerical location of the bit flip, with $000$ meaning no error. Thus, any one bit flip error can be corrected by this code. Up to two bit flip errors can be detected by this code as well, since it can be checked that any two valid code messages are different in at least three bit positions. Other Hamming codes work in a similar fashion. Note that the three parity checks can have one of $2^3$ results, each one giving the location of one of $2^3-1$ bit flip errors or the possibility of no error, making the Hamming codes extremely efficient. There is a precise definition of this in the next section.

\section{Independence Numbers}

An independent set of a graph is a subset of vertices with no edge between them; equivalently, it is a code on the channel defined by the graph. A larger code uses a channel more efficiently, so bounds on the size of independent sets give a better sense of how effective a code is. With no characterization of the errors to be corrected, there can be no nontrivial bounds on code size. The easiest way to measure the error space is with valence. The valence of a vertex is the number of edges containing that vertex. In a regular graph $G$, where each vertex has the same valence, the graph itself is said to have that valence $\size{G}$; in other cases, there is still a definition of minimum valence $\size{G}_{\min}$ and of maximum valence $\size{G}_{\max}$ among the set of vertices. Another useful quantity is the order $\order{G}$ of the graph, which is the number of vertices or channel size. The independence number $\alpha(G)$ is defined as the size of the largest independent set of $G$. Clearly, the independence number should increase with larger channel size and decrease with larger valence. 

\begin{theorem*}For any classical graph $G$, $$\order{G}/(\size{G}_{\max}^2+1)\leq\alpha(G^2)\leq\order{G}/(\size{G}_{\min}+1)$$\end{theorem*}

Recall that error correcting codes on a graph are the same as error detecting codes on the graph's square, so $\alpha(G^2)$ is the maximum size of an error correcting code on $G$. One notable bound on the size of error correcting codes is the volume bound, which says that no error correcting code can have size larger than $\order{G}/(\size{G}_{\min}+1)$. This bound is easily obtained by considering the number of vertices in each neighborhood. If each vertex has a neighborhood of size at least $\size{G}_{\min}+1$, then a code can have size at most $\order{G}/(\size{G}_{\min}+1)$, since all vertices in the code must have disjoint neighborhoods. This is also known as the packing bound, since it is a bound on the number of neighborhoods that can be packed into the graph. Codes that meet this bound are called perfect codes, and the Hamming code is an example of a perfect code. The seven bit channel has $2^7$ vertices, and each vertex has valence $7$, corresponding to single bit flips in each position or no error. Then the volume bound gives a maximum error correcting code size of $2^7/(7+1)=2^4$, which is the number of messages that can be encoded in the four bits that can be transmitted by the $(7,4)$ Hamming code. By a similar calculation, the $(3,1)$ repetition code, which is also the $(3,1)$ Hamming code, is perfect as well. Since $\order{G^2}=\order{G}$ and $$\size{G^2}_{\max}\leq\size{G}_{\max}^2$$ by a simple edge count, the covering bound described below can be translated into the lower bound given above for error correcting codes. Notice that I've chosen to state the looser but asymptotically equal bound.

\begin{theorem*}[Tur\'{a}n]For any classical graph $G$, $$\order{G}/(\size{G}_{\max}+1)\leq\alpha(G)$$\end{theorem*}

This is called the covering bound. The covering bound guarantees that there always exists an error detecting code of size at least $\order{G}/(\size{G}_{\max}+1)$. This is accomplished by a greedy selection algorithm. If a vertex is an element of an independent set, then none of its neighbors can be an element of that set. Iteratively selecting vertices that are not neighbors of previously selected vertices will give an independent set of size at least $\order{G}/(\size{G}_{\max}+1)$, since each element eliminates a neighborhood of at most $\size{G}_{\max}+1$ vertices. The sharpness of this bound is proven by Tur\'{a}n's theorem \cite{PT} applied to the complement graph, and the graphs demonstrating the sharpness of the bound are exactly the complements of the Tur\'{a}n graphs--they are unions of disjoint cliques of the same size.

\section{Shannon Capacities}

Classical channels can often be used multiple times, or multiple copies of the same channel can be used together. Codes that span multiple copies of a channel can often yield better results than codes on individual uses. Channels being used in conjunction are called a product channel. This was used implicitly already in the definition of the Hamming spaces, since strings of bits are products of single bit channels. Conveniently, graph theory has the language to describe the graphs of product channels as a combination of the graphs of each channel in the product. The product graph of two graphs is the graph whose vertex set is the Cartesian product of their vertex sets. In the strong graph product, two pairs in the Cartesian product index adjacent vertices if both components of each are either equal or adjacent, except when they are both equal. If self loops are counted as adjacencies, then the definition simplifies to both components being adjacent. The confusability graph of a product channel is therefore the strong product of confusability graphs, since each copy of the channel is subject to errors independently. The utility of product channels is that they can have codes that are larger than the product of codes on individual channels, where the products of codes are Cartesian products of their elements. The upper bound on the efficiency of such multi-channel codes is called the Shannon capacity. In fact, the Shannon capacity represents the theoretical limit of communication on a noisy channel. The Shannon capacity $\Theta(G)$ of a graph $G$ is defined as $$\Theta(G):=\limsup_{k\to\infty}\sqrt[k]{\alpha(G^{\times k})},$$ where $G^{\times k}$ is the $k$th strong product of $G$ with itself. Since the sequence is bounded by $\order{G}$, the $\limsup$ must exist. It is known that some graphs never attain their Shannon capacity, and all graphs for which Shannon capacity is known attain it with codes on one or two copies or never attain it at all.

\begin{theorem*}For any classical graph $G$, $$\alpha(G)\leq\sqrt[k]{\alpha(G^{\times k})}\leq\Theta(G)$$\end{theorem*}

One  lower bound on the Shannon capacity of a graph is its independence number. This is a result of the fact that the Cartesian products of codes are codes of the strong products of the corresponding graphs. Given a graph $G$ and a largest code $C$, which has cardinality $\alpha(G)$, $C^{\times k}$ is a code of $G^{\times k}$, and $\order{C^{\times k}}=\alpha(G)^k$. Then the independence number of $G^{\times k}$ is always bounded below by $\alpha(G)^k$, and its Shannon capacity must therefore be at least $\alpha(G)$. Code products also show that the sequence $\log(\alpha(G^{\times k}))$ is subadditive, so the sequence $\sqrt[k]{\alpha(G^{\times k})}$ must converge by Fekete's Lemma. Then the $\limsup$ in the definition can be replaced by $\lim$. Also note that, since the independence number bounds the Shannon capacity from below, and the covering bound gives a lower bound on independence number, then the Shannon capacity is also at least $\order{G}/(\size{G}_{\max}+1)$. Conveniently, $\order{G^{\times k}}=\order{G}^k$ and $\size{G^{\times k}}_{\max}+1=(\size{G}_{\max}+1)^k$, so the covering bound scales appropriately with strong products.

\begin{theorem*}[Lov\'{a}sz]For any classical graph $G$, $$\Theta(G)\leq\vartheta(G)$$\end{theorem*}

Not only is the Shannon capacity hard to find, its computational complexity is still unknown. Computing maximum independent sets is NP-complete, so it may be that Shannon capacities are also as difficult to compute. Then it is useful to have a bound on Shannon capacity that can be computed in polynomial time. Such a bound exists, and it is called the Lov\'{a}sz number \cite{LL}, denoted $\vartheta(G)$. The Lov\'{a}sz number of a graph is always larger than its Shannon capacity. While the Lov\'{a}sz number of several families of graphs are known, in few cases does it find the Shannon capacity exactly by matching a known lower bound on Shannon capacity. One notable instance of this is for the cycle graph on five vertices, where the Lov\'{a}sz number is $\sqrt{5}$ and a code of size $5$ exists in the twice strong product of the graph. Then the capacity is found to be exactly $\sqrt{5}$.

\section{Chromatic Numbers}

An interesting quantity to consider in any graph is the chromatic number. A coloring of a graph is a partition of its vertices into independent sets; equivalently, it is a set of independent sets such that each vertex of the graph belongs to exactly one of the independent sets. Each independent set of a coloring is called a color. The chromatic number of a graph $G$, denoted $\chi(G)$, is the size of the smallest coloring of $G$. For error correction, colorings of a channel's confusability graph have been useful for finding large independent sets, but graph colorings are interesting to consider regardless.

\begin{theorem*}For any classical graph $G$, $$\omega(G)\leq\vartheta(\bar{G})\leq\chi(G)$$\end{theorem*}

A clique of a graph is a complete subgraph; it is a set of vertices such that there is an edge between any pair of them. Clearly, the vertices in a clique must all belong to different colors, so the chromatic number is bounded from below by the clique number, which is the size of the largest clique and denoted $\omega(G)$. The complement of a graph is the graph on the same vertex set that has an edge between two vertices if and only if the original graph did not have an edge between the two. Then a clique in a graph is clearly an independent set of the graph's complement. Since the Lov\'{a}sz number of a graph is an upper bound on its independence number, the Lov\'{a}sz number of the graph's complement is an upper bound on the graph's clique number. Interestingly, the Lov\'{a}sz number of a graph lies between the clique and the chromatic number of the graph's complement, which are both NP-complete to compute. Then the Lov\'{a}sz number is also a lower bound of the chromatic number, although it is more complex than the clique number. The clique number bound on chromatic number is sometimes sharp, but it can also be very loose in some cases. For example, Mycielski's Theorem constructs graphs with clique number $2$ (triangle free) and arbitrarily high chromatic number.

\begin{theorem*}[Brooks]For any classical graph $G$, $$\chi(G)\leq\size{G}_{\max}+1$$\end{theorem*}

Another bound on chromatic number is the maximum valence plus one. This bound is proved by a greedy coloring. Take the vertices in any order. For each vertex, it is adjacent to at most $\size{G}_{\max}$ previously colored vertices, so at least one of $\size{G}_{\max}+1$ colors will be free to use. Note that this bound can also be obtained from the covering bound. Since each color is an independent set, $\order{G}\leq\alpha(G)\chi(G)$ for any graph $G$. This, combined with the covering bound, gives the $\size{G}_{\max}+1$ bound above. For error correcting codes, the bound can be adapted to $$\size{G}_{\min}+1\leq\chi(G^2)\leq\size{G}_{\max}^2+1$$ when combined with the packing bound as well. The bound in the theorem is sharp, but the chromatic number is bounded by just the maximum valence in most cases. This result is the second part of Brooks' Theorem, which also states that the only graphs that have chromatic number equal to maximum valence plus one are odd cycles and complete graphs. The bound along with the sharpness result are often collectively called Brooks' Theorem, although Brooks's Theorem is the grammatically correct name. Considering odd cycles as a special case in valence $1$, Brooks' Theorem states that only in the worst case, a complete graph, is the last color necessary. Star graphs demonstrate that there are graphs with chromatic number two and arbitrarily high maximum valence, so the bound can be quite loose. The following theorem by Erd\H{o}s \cite{PE} further demonstrates the complexity of chromatic numbers. A tree is a graph without cycles, and the chromatic number of any tree is two by a greedy coloring. The girth of a graph is the length of its shortest cycle, so graphs of high girth locally resemble trees. However, the theorem proves that there exist graphs of arbitrarily high girth and chromatic number.

\chapter{General Channels}

Classical and quantum error correction are very similar because they are, in fact, just two sides of the von Neumann approach to information theory. There are three equivalent definitions of a von Neumann algebra, and each has its benefits in describing general systems. As in the classical case, channels can be given as maps, as relations, or as graphs. Similar error correction can be described, of which the previous chapter is a special case. In the next chapter, another case is considered: the fully quantum case.

\section{As Maps}

A $^*$-algebra $A$ is defined as a Banach algebra over $\mathbb{C}$ with an antilinear, antiautomorphic involution $^*$. In other words, it is an algebra over $\mathbb{C}$ with the following properties: \begin{align*}x^{**}&=x\\(xy)^*&=y^*x^*\\(\lambda x)^*&=\bar{\lambda}x^*\\(x+y)^*&=x^*+y^*\end{align*} for all $x,y\in A$ and $\lambda\in\mathbb{C}$. A C$^*$-algebra is a $^*$-algebra with an added property: $$\size{x^*x}=\size{x^*}\size{x}$$ for all $x\in A$. The importance of C$^*$-algebras is that they describe algebras of operators, with $^*$ being the adjoint operation. Not all C$^*$-algebras will describe a complete set of operators of a quantum system, however; another condition is required. A W$^*$-algebra is a C$^*$-algebra with a predual as a Banach space, or, in other words, there exists a Banach space such that its dual is the W$^*$-algebra.

\begin{definition*}A general channel is a TPCP (trace preserving, completely positive) map on the predual of a W$^*$-algebra \cite{SS}.\end{definition*}

Since each W$^*$-algebra has a predual that is unique up to isomorphism, this definition is unambiguous. A classical channel was defined as a matrix of probabilities, but it is also a linear map with two defining properties. The first property is that it conserves probability, so it preserves the norm of each state. The second property is positivity, which is necessary to keep probabilities nonnegative. A general channel is defined similarly as a linear map on the predual of a W$^*$-algebra that is trace preserving and completely positive. Preserving trace means that it preserves the Hilbert-Schmidt norm. Preserving positivity takes more work. A positive element of the predual is defined as an element that can be represented as $x^*x$ for some element $x$. A positive map is one that takes positive elements to positive elements. Two maps that are positive on preduals $A$ and $B$, however, may not combine into a positive map on $A\otimes B$. This is unfortunate because the W$^*$-algebra of a joint quantum system is the tensor product of the W$^*$-algebras of its component systems. Then the positive maps should tensor accordingly, which is described by the following property. A map $M$ on a system $A$ is completely positive if $M\otimes I$ is positive on $A\otimes B$ for any auxiliary system $B$. Complete positivity is preserved on joint maps, and so it is the appropriate condition for a channel.

\begin{theorem*}[Murray,von Neumann]Every commutative W$^*$-algebra is isomorphic to $L^\infty(X)$ for some finitely decomposable measure space $X$. Conversely, $L^\infty(X)$ is a W$^*$-algebra for every finitely decomposable measure space $X$ \cite{MN}.\end{theorem*}

General quantum mechanics models a system as a W$^*$-algebra called its von Neumann algebra. A system's observables correspond to the self adjoint ($x=x^*$) elements of its von Neumann algebra, so it can also be called an algebra of observables. Observables are also called measurements, and they represent ways to obtain information about and to condition the states of the system. The states of a system, in this formulation, are the positive ($x=y^*y$ for some $y$) and unital ($\size{x}=1$) elements of the predual of its von Neumann algebra; the results of measurements on states are given naturally by the dual relationship of their Banach spaces. Consider the W$^*$-algebras that are commutative and finite dimensional. By the above theorem, any commutative W$^*$-algebra is isomorphic to some $L^\infty(X)$. The finite dimensional commutative W$^*$-algebras are therefore $L^\infty(X)$ for a finite set $X$ with counting measure. Then the predual is $L^1(X)=l^1(X)$, and its positive and unital elements correspond to the state spaces described in the previous chapter. Note that the trace preserving and completely positive maps on such spaces are exactly stochastic matrices. Thus, classical systems are the fully commutative and finite dimensional general systems in this formulation. In classical probability, the von Neumann algebra of a classical system is the algebra of bounded, complex random variables. In its sigma algebra of events, $+$ and $\times$ correspond to classical XOR and AND.

\section{As Relations}

For any Hilbert space $H$, the space of bounded operators on it form a W$^*$-algebra called $B(H)$ with the $^*$ action given by taking adjoints. An operator in $B(H)$ is trace class if its trace can be defined, meaning that its trace is finite and independent of the choice of basis. The predual of $B(H)$ is called $B_t(H)$, and it consists of all trace class operators with norm given by the trace; its action on $B(H)$ is given by $$A(B)=\operatorname{tr}(AB).$$ The weak$^*$ topology can be defined on $B(H)$ as the weakest topology in which the trace class operators act continuously on $B(H)$. This leads to another characterization of W$^*$-algebras--as C$^*$-subalgebras of $B(H)$ that contain the identity operator and are closed in the weak$^*$ topology.

\begin{definition*}A general relation on a W$^*$-algebra $M\subseteq B(H)$ is a weak* closed subspace $R$ of $B(H)$ such that $M'RM'\subseteq R$, where $M'$ is the commutant of $M$ \cite{NW}.\end{definition*}

As in the classical case, a general channel can be reduced to a general relation between states. In fact, a general relation on $M=l^\infty(X)$ is equivalent to a classical relation on $X$ for any finite set $X$, although the equivalence is not immediate. A general relation generalizes the matrix form of a classical relation, for which there is a $0$ in the $i,j$th position for every $i$ and $j$ not related. The subspace of all matrices with the appropriate zero entries is the corresponding general relation. The commutant will be discussed below, but representations must first be discussed. A general relation is defined here only on W$^*$-algebras that are subalgebras of $B(H)$ for some $H$. In fact, a relation can be defined on any W$^*$-algebra. A homomorphism of a W$^*$-algebra into $B(H)$ is called a representation, and it is faithful if it is injective. By the theorem below, any W$^*$-algebra has a faithful representation in some $B(H)$, so a general relation can be defined on such a representation. Given two faithful representations of a W$^*$-algebra, there is a one-to-one correspondence between their general relations, so this definition is independent of the representation chosen.

\begin{theorem*}[Gelfand,Naimark]Every W$^*$-algebra is a subalgebra of $B(H)$ for some Hilbert space $H$ \cite{GN}.\end{theorem*}

An element $P$ of a W$^*$-algebra is called a projection if it is self adjoint ($P=P^*$) and idempotent ($P=P^2$); note that these are typically called orthogonal projections. The only projections in $B(H)$ that are states are those that have rank one, and they are called pure states; the space of pure states is isomorphic to the space of unital elements of $H$ excluding global phase. An arbitrary projection in $B(H)$ will correspond to a subspace of $H$, so a general code is defined as a projection much as a classical code is defined as a subset of pure states. Given a general relation $R$ on a W$^*$-algebra $M\subseteq B(H)$, the error detection condition that a code $P$ must satisfy is $$PEP=\epsilon(E)P$$ for any $E\in R$ and some function $\epsilon:R\to M'$. The epsilon function is necessary as a result of the condition $M'RM'\subseteq R$; as will be discussed below, the commutant has no effect on measurements. The error detection procedure is simple. If an error $E$ occurs on $P$, then measure the resulting condition with $P$. The result is either $P$, in which case the state is recovered, or it is orthogonal to $P$, and an error is detected but cannot be corrected. Similar to the classical case, a code detecting errors in $R^2$ can correct errors in $R$ for any general relation $R$, but this will be discussed more in the next chapter.

\section{As Graphs}

The commutant $M'$ of a W$^*$-algebra $M\subseteq B(H)$ is the algebra of elements in $B(H)$ that commute with every element of $M$. The commutant of a W$^*$-algebra represents the excess of operators in an embedding; if a larger than necessary Hilbert space is chosen, the difference will be reflected in the commutant. It can also be viewed as the algebra of unobservables in that any element commuting with all observables will have no effect on the measurement of any observable and so cannot be observed in the system. In light of this fact, any actions of the commutant should not affect relations between elements either, so, to define a general relation $R$, $$M'RM'\subseteq R$$ or, equivalently, $$M'RM'=R,$$ since the commutant always includes the identity. It is this condition that allows general relations to be defined independently of representation. A third characterization of W$^*$-algebras is as C$^*$-algebras that are equal to their double commutants: $M=M''$.

\begin{definition*}A general graph is a reflexive and symmetric general relation; in other words, it is a weak$^*$ closed operator system that is a bimodule over the commutant \cite{KW}.\end{definition*}

Reflexivity of a relation is simple to define. Since the identity operator relates every element to itself, a general relation containing the identity is called reflexive. This also implies that the relation contains all of the commutant $M'$, since $M'RM'$ would then contain $M'$. A general relation is called symmetric if it is closed under $^*$. This is because the adjoint of an operator gives the reverse relation. Since $$PE^*P=(PEP)^*$$ implies $$\epsilon(E^*)=\epsilon(E)^*,$$ it is clear that error detecting codes will, as in the classical case, detect neither or both directions of a relation. An operator system is defined as a subspace of operators closed under $^*$ and containing the identity, so it is exactly the definition of a quantum graph when combined with the bimodule over commutant condition carried over from the definition of a general relation. The vertex space of a general graph is its Hilbert space. Two code projections $P$ and $Q$ are not independent if and only if $$PEQ\neq 0$$ for some $E$ in the general graph; this is a generalization of a classical independence, since projections onto the subspaces spanned by independent sets of classical pure states will satisfy the independence condition above. A coloring of a general graph is a partition of unity as a sum of disjoint projections, so it is a set of codes which sum to the identity. Independent sets and colorings will be discussed more in the next two chapters.

\begin{theorem*}[Artin,Wedderburn]Every finite dimensional W$^*$-algebra is isomorphic to a sum of matrix algebras: $\bigoplus_k M_{n_k}$ for some sequence $n_k$ \cite{JW}.\end{theorem*}

In finite dimensions, the picture is clear. A finite dimensional W$^*$-algebra is the commutative sum of noncommutative matrix algebras. Thus, the most commutative ones are direct sums of dimension $1$ matrix algebras, $\bigoplus_k M_1=\mathbb{C}^k=l^\infty(k)$. The least commutative ones are single matrix algebras, $M_n=B(\mathbb{C}^n)$. In this formulation, classical and quantum systems are systems that have the most commutative and least commutative von Neumann algebras respectively. The commutativity of an arbitrary W$^*$-algebra can be determined by the size of its center. If a W$^*$-algebra is classical, then its center is itself. If a W$^*$-algebra is quantum, then its center is as small as possible; in other words, it is the set of all multiples of the identity. If the center is anywhere in between, then the algebra is neither fully classical nor fully quantum. Classical (commutative) W$^*$-algebras are characterized above, but quantum W$^*$-algebras are more complex. The simplest type of quantum W$^*$-algebras are $B(H)$ for a Hilbert space $H$. However, there are other W$^*$-algebras that also have trivial centers, and these are called factors. Factors are classified as one of three types, and $B(H)$ is of type 1. In the finite dimensional case, however, all factors are of type 1, so those are the ones discussed here. 

\chapter{Quantum Channels}

All of the necessary terminology and concepts have been laid out, in parallel, in the first chapter. As with classical channels, quantum channels can be converted from maps to relations to graphs. There are, as well, the concepts of states, error correcting codes, and error detecting codes, and examples are given of each. For the study of quantum graphs, three graph invariants are introduced: independence number, Shannon capacity, and chromatic number. A bound is given for each invariant, but it is clear that the classical case has seen far more study; this is partially remedied in the next chapter.

\section{As Maps}

If a Hilbert space $H$ is finite dimensional, then $B(H)$ is a matrix algebra, and a fully quantum system has that whole matrix algebra as its von Neumann algebra. As defined in the previous chapter, a channel is a trace preserving and completely positive map. It follows from the proof of Choi's theorem \cite{MC} that a completely positive linear operator $M$ on a matrix algebra is given by a set of matrices $\{M_i\}$ with action $$M(A)=\sum M_iAM_i^*.$$ It also follows by a dimension count that no more than $\dim(H)^2$ matrices are required to describe such a map. This is called the operator sum representation of a completely positive linear map, and its matrices are also called Kraus operators. The choice of Kraus operators is not unique. Two sets of matrices will represent the same operator if they are related in a certain way by a unitary matrix. Given a completely positive map represented by $\{M_i\}$, it is also trace preserving if $$\sum M_i^*M_i=I.$$ Since a pure state is $|a\rangle\langle a|$ for some $a\in H$, the action $$\sum M_i|a\rangle\langle a|M_i^*=\sum(M_i|a\rangle)(M_i|a\rangle)^*$$ is also an action on the Hilbert space corresponding to the pure states. The transition probability $p(a,b)$ between two pure states $a,b\in H$ can also be defined from a quantum channel: $$p(a,b)=\sum\order{\braket{a|M_i|b}}^2.$$ This formula can be obtained by taking the operator corresponding to $a$, applying the channel operator, and measuring the result with $b$.

\begin{definition*}A quantum channel is a set of square matrices $\{M_i\}$ such that $\sum M_i^*M_i=I$.\end{definition*}

For a fully quantum von Neumann algebra, its set of states consists of matrices that are self adjoint with nonnegative eigenvalues summing, with multiplicity, to $1$. These are called density matrices, and they represent classical mixtures of pure quantum states. The pure states of this set, as in the classical case, are the extremal points. It is clear that the extremal points of the set of density matrices are the rank $1$ matrices, and these can each be expressed as $|a\rangle\langle a|$ for some element $a\in H$, the base Hilbert space. Thus, the space of pure states can, as before, be identified with the unital elements of $H$ excluding global phase. As with the classical case, the action of the channel on arbitrary states is determined by its action on the pure states and extending linearly. A map that can be represented by a single Kraus operator is called coherent. It follows that any completely positive, trace preserving, and coherent map can be represented by a single unitary matrix, and this also characterizes all maps that take pure states to pure states. These correspond to the unitary operators of the less complete Hilbert space model of states. The linearity of such maps is called quantum superposition. Along with classical superposition, it implies the linearity of fully general channels.

\begin{definition*}A quantum state is a positive matrix with trace $1$.\end{definition*}

A quantum Hamming space is an example of a quantum channel. A qubit is a unit of quantum information, and its Hilbert space is $\mathbb{C}^2$. Then its von Neumann algebra is the matrix algebra $M_2(\mathbb{C})$. The Hilbert space $\mathbb{C}^2$ has orthonormal basis elements commonly called $0$ and $1$. The algebra $M_2(\mathbb{C})$ has a commonly used basis called the Pauli matrices; it consists of the matrices $$I=\begin{pmatrix}1&0\\0&1\end{pmatrix}, X=\begin{pmatrix}0&1\\1&0\end{pmatrix}, Y=\begin{pmatrix}0&-i\\i&0\end{pmatrix}, Z=\begin{pmatrix}1&0\\0&-1\end{pmatrix}.$$ The matrix $X$ is also called the bit flip error, since its effect on a qubit is to exchange the $0$ and $1$ states. The matrix $Z$ is also called the sign flip error because it exchanges the $-$ and $+$ states. A quantum Hamming space is a product of qubits. If it is a product of $n$ qubits, its Hilbert space and von Neumann algebra are $(\mathbb{C}^2)^{\otimes n}=\mathbb{C}^{2^n}$ and $M_2(\mathbb{C})^{\otimes n}=M_{2^n}(\mathbb{C})$, respectively. Errors on a quantum Hamming space are products of arbitrary errors on individual qubits. An error on $k$ qubits therefore has the form $E_1\otimes\ldots\otimes E_n$ where each $E_i\in M_2(\mathbb{C})$ and all but $k$ of the $E_i$ are the identity operator. The error space of arbitrary errors on at most $k$ qubits, then, has a basis of the form $$\{E_1\otimes\ldots\otimes E_n:E_i\text{ is a Pauli matrix for all }i\text{, and all but at most }k\text{ of the }E_i\text{ are the identity}\}.$$

\section{As Relations}

The general definition of a relation becomes simpler when restricted to fully quantum channels. If the von Neumann algebra is all of $B(H)$, then a relation is a dual operator space: it is a weak$^*$ closed subspace of $B(H)$. Furthermore, if $H$ is finite dimensional, then the weak$^*$ closure property holds automatically, and a relation is simply a subspace of $B(H)$. As in the classical case, the relation is determined by its action on pure states, and two pure states $a,b\in H$ are related by a quantum relation $R\subseteq B(H)$ if and only if $$\braket{a|E|b}\neq0$$ for some $E\in R$. Note that the condition $\braket{a|E|b}=0$ extends to linear combinations of matrices in $R$, so considering only a basis of $R$ is equivalent to considering all of $R$. From the transition probability defined above, this exactly corresponds to rounding small probabilities to $0$, since $\braket{a|E|b}=0$ for all $E\in R$ if and only if $$\sum_E\order{\braket{a|E|b}}^2=0$$ for a sum over any basis of $R$. Then reducing a channel to a relation is equivalent to choosing a subset of Kraus operators in some representation and taking the subspace it spans as the relation.

\begin{definition*}A quantum relation is a subspace of square matrices.\end{definition*}

A quantum error correcting code is a subspace of pure states such that any two orthogonal elements of the subspace are still orthogonal when each is acted upon by an error. If the subspace is $C$, then this means $\braket{a|E^*F|b}=0$ for any orthogonal $a,b\in C$ and $E,F\in R$. This is equivalent to saying that $C$ detects errors from $R^*R$, and it will be discussed in the next section. Unlike the classical case, the neighborhoods of any two orthogonal elements in an error correcting code may not be disjoint; a code for which such neighborhoods are always disjoint is called a nondegenerate code. For nondegenerate codes, there is an embedding of $R\otimes C$ into $H$ by matrix multiplication. For degenerate codes, more work is required to achieve such an embedding. Let $P$ denote the projection onto $C$. Recall from the previous chapter the condition $PEP=\epsilon(E)P$ for error detection, which is equivalent to the condition $P(E^*F)P=\epsilon(E^*F)P$ for error correction; in the fully quantum case, $\epsilon$ is a scalar function. Then $\braket{E,F}=\epsilon(E^*F)$ is a Hermitian form on $R$ which is degenerate if and only if the code $C$ is degenerate. The kernel of this Hermitian form is the overlap in neighborhoods, so the quotient of $R$ by the kernel, call it $R_0$, gives an embedding $R_0\otimes C$ into $H$. This proves the quantum Hamming bound, which is that $$\dim(R_0)\dim(C)\le\dim(H)$$ for any quantum relation $R$ and any error correcting code $C$ of the relation. Nondegenerate codes for which this bound is sharp are called perfect quantum codes. For the $n$ qubit Hamming space described above and arbitrary $1$ qubit errors, the inequality becomes $$(3n+1)\dim(C)\le 2^n.$$ For the transmission of $1$ qubit, $\dim(C)=2$, and the smallest value of $n$ for which this bound holds is $5$. In fact, the bound is sharp for $n=5$. There exists an error correcting code encoding $1$ qubit into $5$, and so it is perfect.

\begin{definition*}A quantum error correcting code is a subspace $C$ of a Hilbert space $H$ such that $\braket{a|E^*F|a}=\epsilon(E,F)\braket{a|a}$ for any $E,F\in R$ and any $a\in C$.\end{definition*}

An example of an error correcting code is the Shor code \cite{PS}, which is related to the classical repetition code. It encodes $1$ qubit into $9$, so it is a $(9,1)$ quantum code. The errors it corrects are the arbitrary single qubit errors which were described in the previous section. Denote the tensor $a_1\otimes\ldots\otimes a_n$ by $a_1\ldots a_n$. Note first that the $3$ qubit code $$\operatorname{span}(\{000,111\})$$ can correct single bit flip errors. It can do this by measuring an errored state by $$\{\operatorname{span}(\{000,111\}),\operatorname{span}(\{100,011\}),\operatorname{span}(\{010,101\}),\operatorname{span}(\{001,110\})\}$$ to determine which qubit incurred the error. Denote the state $(0+1)/\sqrt{2}$ by $+$ and the state $(0-1)/\sqrt{2}$ by $-$. Then also note that the $3$ qubit code $$\operatorname{span}(\{+++,---\})$$ can correct single sign flip errors by a similar measurement because sign flip errors exchange $+$ and $-$ just as bit flip errors exchange $0$ and $1$. Then the product of these two codes, \begin{align*}\operatorname{span}(\{&((000+111)\otimes(000+111)\otimes(000+111))/(2\sqrt{2}),\\&((000-111)\otimes(000-111)\otimes(000-111))/(2\sqrt{2})\})\end{align*}, can correct a single bit flip, sign flip, or both by the appropriate measurement. Both errors occurring is equivalent to a $Y$ error because $XZ=iY$, so the whole Pauli basis can be corrected by the Shor code, and, by extension, all arbitrary single qubits errors can be corrected. The Shor code is an example of an additive code, which is also called a stabilizer code. Define a multi-Pauli as a tensor product of Pauli matrices. The set of multi-Paulis of a fixed dimension forms a group by multiplication, and every abelian subgroup corresponds to a space of states that are unaffected by every element of that subgroup. Then that space of states is a code that corrects against every multi-Pauli in the subgroup as well as any multi-Pauli that anticommutes with some element of the subgroup, since measuring with those elements will identify the error. Codes defined in this way are called additive codes. For example, the two subgroups $$\{III,IZZ,ZIZ,ZZI\}$$ and $$\{III,IXX,XIX,XXI\}$$ will stabilize the two $3$ qubit codes described above.

\section{As Graphs}

Classical graphs are equivalent to reflexive and symmetric classical relations. Similarly, quantum graphs are equivalent to reflexive and symmetric quantum relations, where a quantum relation is reflexive if it contains the identity and symmetric if it is closed under $^*$. To copy the language of classical graph theory, the vertices of a quantum graph are its pure states, so the space of vertices $V(G)$ of a quantum graph $G$ is $H$, the Hilbert space of pure states. The subspace that defines the quantum graph is the space of edges $E(G)$, and the inclusion of the identity means that self loops are also included. A notable difference between the definition of classical and quantum graphs is that quantum graphs are defined as the set of edges acting on each vertex, unlike classical graphs, which are the sets of all edges acting on any vertex. In consequence, all quantum graphs are regular, since each vertex has the same edge space and therefore the same valence.

\begin{definition*}A quantum graph is a subspace of square matrices closed under $^*$ and containing the identity.\end{definition*}

The general definition of an error detecting code is a projection $P$ that satisfies the error detecting condition: $$PEP=\epsilon(E)P$$ for every $E$ in a general graph $G$ \cite{KL}. For quantum graphs, a projection in $B(H)$ will correspond to a subspace of $H$. Then an equivalent condition in the quantum case for a subspace $C$ to be an error detecting code is that $$\braket{a|E|a}=\epsilon(E)\braket{a|a}$$ for every vector $a\in C$ and every $E$ in a quantum graph $G$. Note that this condition extends to linear combinations of errors, so only a basis of the error space needs to be detectable for the whole space to be. Furthermore, every quantum graph has a basis of self adjoint matrices, since the space is closed under $^*$. This can be seen by taking an arbitrary basis $\{B_i\}$ and choosing a new basis out of the set $\{B_i+B_i^*,i(B_i-B_i^*)\}$ of self adjoint matrices. Moreover, the condition also extends to linear combinations of states as well given that $\braket{a|E|b}=0$ for any orthogonal vectors $a,b$ in the code and all $E$ in the quantum graph. Therefore, a necessary and sufficient condition for a set of orthogonal vectors to be a basis of an error detecting code is that $$\braket{a|E|b}=\epsilon(E)\braket{a|b}\delta_{a,b}$$ for any vectors $a,b$ in the set and any $E$ in the quantum graph. An independent set can be defined as a set $\{a_i\}$ of orthogonal vectors such that they are orthogonal even under one error: $$\braket{a_i|E|a_j}=0$$ for all $E$ in the quantum graph if $i\neq j$. However, since any orthogonal basis of an error detecting code will satisfy this condition and any such basis will span an error detecting code, an independent set can more appropriately be defined as an error detecting code.

\begin{definition*}A quantum error detecting code is a subspace $C$ of a Hilbert space $H$ such that $\braket{a|E|a}=\epsilon(E)\braket{a|a}$ for any $E\in G$ and $a\in C$.\end{definition*}

Another example of a quantum code is the Steane code. The Steane code is an additive code, and it is stabilized by the subgroup generated by the set $$\{IIIXXXX,IXXIIXX,XIXIXIX,IIIZZZZ,IZZIIZZ,ZIZIZIZ\}.$$ The Steane code can also be constructed as a CSS code from the classical Hamming code in a way not very different from the construction of the Shor code from the classical repetition code. More importantly, it can demonstrate how a coloring can be obtained from any additive code. Take any additive code and apply every possible multi-Pauli. Clearly, this will result in a set of subspaces with the same dimension and coloring the space. To show that each subspace is a code correcting the same errors, consider the eigenvalues of the code elements on each multi-Pauli in the generating set. The Steane code itself will have eigenvalue $1$ on each multi-Pauli. The Steane code shifted by a multi-Pauli will have eigenvalue $-1$ on some subset of the generating set, and every code element will have the same eigenvalues. In fact, each distinct subspace will have eigenvalue $-1$ on a different subset of generating elements, and each subset will correspond to a subspace. That each subset can be separated in this way is a consequence of the generating set being minimal, and each element having order $2$ implies that each minimal generating set has the same size. The multi-Pauli shifting the code can then be identified by which subset of generators it anticommutes with, and the same errors can be corrected. There are $6$ generators, so there are $2^6$ subsets, each corresponding to a one qubit code. Since each qubit has dimension $2$, this perfectly partitions the space of dimension $2^7$. While colorings can be constructed this way, this construction only works for the quantum Hamming space, so it does not apply to quantum graphs in general.

\section{Graph Invariants}

The independence number $\alpha(G)$ of a quantum graph $G$ is defined as the dimension of its largest code. One less than the dimension of the edge space of a quantum graph $G$ is called its valence, and it is denoted by $\size{G}$. The reason for subtracting one is to exclude the identity from the valence count, just as self loops are excluded from classical valence counts. The dimension of the vertex space is denoted by $\order{G}$. There is one bound on quantum independence numbers \cite{KLV}, and it is similar to the covering bound for classical independence numbers. The first half of the proof, in fact, is the same greedy algorithm as in the proof of the covering bound. The second half of the proof is a clever application of Tverberg's theorem. In each step, a factor of $1/(\size{G}+1)$ is introduced, so the end result is a lower bound of $\order{G}/(\size{G}+1)^2$. The full details of the proof are provided in the next chapter.

\begin{theorem*}For any quantum graph $G$, $$\order{G}/(\size{G}+1)^2\le\alpha(G)$$\end{theorem*}

The product of two quantum graphs is their tensor product. This definition is a little simpler than in the classical case since self loops are included in the edge space, as is natural. The Shannon capacity of a quantum graph can then be defined in exactly the same way as that of a classical graph. A Lov\'{a}sz number can be defined for a quantum graph as well, and it once again bounds the Shannon capacity from above. There is a full treatment of this \cite{DSW}, in which the Shannon capacity is called the entanglement assisted zero error capacity.

\begin{theorem*}For any quantum graph $G$, $$\chi(G)\le\slog(\order{G},(\size{G}+1)^2)$$\end{theorem*}

The chromatic number $\chi(G)$ of a quantum graph $G$ is defined as the cardinality of its smallest coloring. A coloring is a set of error detecting codes which sum directly to $H$. Note that each code is allowed to have different values for $\epsilon$. Recall that colorings can be obtained from additive codes \cite{CRSS}, but this method only applies to one family of quantum graphs. The only bound on the chromatic numbers of all quantum graphs is one that is derived from the bound on independence number. The chromatic number bound can be obtained by iteratively taking codes of the size guaranteed by the independence number bound. This quantity is asymptotically a logarithm, but its exact value can be computed. I call it the step logarithm, and the next chapter contains a discussion of the $\slog$ function. It should be noted that, in writing this paper, the author has become quite familiar with the definition of slog. It is clear that there is a shortage of bounds on quantum graph invariants, and particularly notably absent is a quantum analog of Brooks' theorem. That is, there is no upper bound on the chromatic numbers of quantum graphs that depends only on valence. In the next chapter, I prove an upper bound on independence number and a lower bound on chromatic number opposite of the bounds provided here and, as a consequence, show that no such Brooks' bound can exist for quantum graphs.

\chapter{My Results}

Recall the following definitions from the previous chapter. A \emph{quantum graph} $G$ is a pair of a finite dimensional complex inner product space $V(G)$ and a real vector space $E(G)$ of self adjoint operators on $V(G)$ that contains the identity. Denote $\order{G}:=\dim(V(G))$ and $\size{G}:=\dim(E(G))-1$. A \emph{code} $C$ of $G$ is a subspace of $V(G)$ such that $P_CAP_C=\epsilon_C(A)P_C$ for all $A\in E(G)$ and some $\epsilon_C:E(G)\to\mathbb{R}$, where $P_C$ is the projection onto $C$. A \emph{coloring} $K$ of $G$ is a set of codes of $G$ such that $\sum_{C\in K}P_C=I$.

\section{Slopes/Eigenvalues}

The first two lemmas establish links between the sets of slopes and eigenvalues. The first concerns medians, and the second concerns means. The incompatibility of these two lemmas results in the main theorems of this paper. The function $\epsilon_C$ is called the code's \emph{slope}. Note that for a single eigenvector $v$ of a matrix $A$, $\epsilon_{\operatorname{span}(v)}(A)$ is just the eigenvalue corresponding to $v$, so the slope can be seen as one generalization of the eigenvalue. Given a code $C$ of a quantum graph $G$, there is a basis of $E(G)$ consisting only of the identity and operators with slope $0$. It can be obtained by taking an arbitrary basis containing the identity and substituting $A-\epsilon_C(A)I$ for each other element $A$. Then, among elements $x,y$ of the code, $\braket{x|A|y}=0$ for every element $A$ of this basis that is not the identity. Alternatively, $\braket{x|A|y}=0$ for all $A$ and any orthogonal elements $x,y$ of the code. Then the code can be seen as containing elements with no other edges between them beside self loops, as in the classical case.

\begin{lemma}\label{isotropic}If $C$ is a code of a quantum graph $G$, the dimension of $C$ is at most equal to the cardinality of $\sigma(A)_{\ge\epsilon_C(A)}$ for any $A\in E(G)$.\end{lemma}

\begin{proof}For any $A\in E(G)$, $C$ is totally isotropic with respect to the Hermitian form given by $A-\epsilon_C(A)I$. The dimension of a totally isotropic subspace is at most equal to the minimum of the number of nonpositive eigenvalues and the number of nonnegative eigenvalues, so it is at most equal to the number of nonnegative eigenvalues in particular.\end{proof}

\begin{lemma}\label{trace}If $K$ is a coloring of a quantum graph $G$, the multisets $\sigma(A)$ and $\epsilon_K(A)$(with multiplicity by dimension) have the same mean for any $A\in E(G)$.\end{lemma}

\begin{proof}Take an orthonormal basis of each code $C\in K$, and combine them into an orthonormal basis of $V(G)$. For any $A\in E(G)$, the sum of the multiset $\epsilon_K(A)$ is the trace of $A$ expressed in the constructed basis, so it is equal to the sum of the multiset $\sigma(A)$. Since both multisets have cardinality $\order{G}$, they have the same mean.\end{proof}

\section{Tropicality/Cyclicality}

The lemmas above are suggestive of a worst case. A construction is chosen to most effectively separate the median and the mean of the sets of eigenvalues. Then, the next two lemmas establish that much of this separation is preserved even when restricted to subspaces of this construction. Let $Q_{n,m}$ be a quantum graph such that $\order{Q_{n,m}}=n$ and $\size{Q_{n,m}}=m$, and $E(Q_{n,m})$ is \emph{commutative}, meaning that 
$$AB=BA$$
 for any $A,B$ in $E(Q_{n,m})$, and has a basis $\{I,A_i\}$ that is \emph{tropical}, meaning that 
$$0<\lambda_{i,j+1}<\lambda_{i,j}/n^2$$
 for all $i,j$, and \emph{cyclical}, meaning that 
$$w_{i+1,j}=w_{i,j+\floor{n/m}+\delta_{[1,n\mod m]}(i)\mod n}$$
 for all $i,j$, where $\lambda_{i,j}$ and $w_{i,j}$ are the $j$th eigenvalue and eigenvector respectively of $A_i$. The tropical condition just means that the eigenvalues of each matrix are numbered in decreasing order and are a certain degree of magnitude apart; note that the fully tropical limit is not required here, and the chosen magnitude is sufficient. The cyclical condition just means that the eigenvectors of the matrices, which can all match because of commutativity, are numbered in cyclic shifts such that they are as evenly distributed as possible; in the case that $m|n$, the condition is simply 
$$w_{i+1,j}=w_{i,j+n/m\mod n}.$$
 Relabel the mutual eigenvectors singly in reverse lexicographic order, so, for all $k$, each vector $w_{(k-1)m+1}$ through $w_{km}$ is equal to $w_{i,k}$ for some $i$. Now, each eigenvector is labelled $m$ times with two indices and $1$ time with one index. Both labelling schemes will be useful in the following proofs. To demonstrate all of this, take $n=10$ and $m=3$. The constructed basis of the edge space would consist of the identity and the following three errors: 
$$0123456789,6789012345,3456789012,$$
 where the eigenvalues of each matrix are ordered from $0$ (highest) to $9$ (lowest) and the place of each digit represents the matrix's value on that shared eigenvector. The relabeling step shuffles the eigenvectors, so the errors are now: 
$$0471582693,6037148259,3704815926.$$
 Notice that the first three eigenvectors each take value $0$ on some error, the next three eigenvectors each take value $1$ on some error, the next three eigenvectors each take value $2$ on some error, and the last eigenvector takes value $3$ on the first error. Not only are the eigenvalues of each matrix distributed tropically, the tropicality is distributed across dimensions as evenly as possible, so that any removal of dimensions can only reduce tropicality by a bounded amount; this is made precise in the following lemmas. The commutativity of an edge space can be interpreted in the following way. Take the edge space, $E(G)$, to the infinite product; this set would consist of all finite products of elements in $E(G)$. The edge space is an operator system, and the infinite product closes it under products, so the result is a W$^*$-algebra that is still commutative. Then it is a classical von Neumann algebra, and its elements are connected components(and mixtures of them). Then commutativity of a quantum graph implies that its connected components interact classically.

\begin{lemma}\label{rayleigh}If $S$ is a subspace of $V(Q_{n,m})$, then $$\lambda_{\max}(P_SA_iP_S)\ge\lambda_{i,j}\braket{w_{i,j}|P_S|w_{i,j}}$$ for any $i,j$.\end{lemma}

\begin{proof}The maximal eigenvalue of a Hermitian matrix $M$ is the largest value achieved by its Rayleigh quotient $$R_M(x)=\braket{x|M|x}/\braket{x|x}.$$ Then, for any $i,j$, \begin{align*}
\lambda_{\max}(P_SA_iP_S)
&\ge\braket{P_Sw_{i,j}|P_SA_iP_S|P_Sw_{i,j}}/\braket{P_Sw_{i,j}|P_Sw_{i,j}}\\
&=\braket{w_{i,j}|P_SA_i(\sum_k|w_{i,k}}\braket{w_{i,k}|)P_S|w_{i,j}}/\braket{w_{i,j}|P_S|w_{i,j}}\\
&=\sum_k\lambda_{i,k}\order{\braket{w_{i,k}|P_S|w_{i,j}}}^2/\braket{w_{i,j}|P_S|w_{i,j}}\\
&\ge\lambda_{i,j}\braket{w_{i,j}|P_S|w_{i,j}}
\end{align*} by the positivity of $A_i$ and $P_S$.\end{proof}

\begin{lemma}\label{basis}If $S$ is a nontrivial subspace of $V(Q_{n,m})$, then $$\braket{w_k|P_S|w_k}\ge1/(\dim(S^\perp)+1)$$ for some $k\le\dim(S^\perp)+1$.\end{lemma}

\begin{proof}Take an orthonormal basis $B_0$ of $$S\cap\operatorname{span}(\{w_k\}_1^{\dim(S^\perp)+1}),$$ and extend it to an orthonormal basis $B_1$ of $S$, and extend it to an orthonormal basis $B_2$ of $V(Q_{n,m})$. Then \begin{align*}
\sum_{k=1}^{\dim(S^\perp)+1}\braket{w_k|P_S|w_k}
&=\sum_{k=1}^{\dim(S^\perp)+1}\sum_{x\in B_2}\braket{w_k|P_S(|x}\braket{x|)w_k}\\
&=\sum_{k=1}^{\dim(S^\perp)+1}\sum_{x\in B_1}\braket{w_k|x}\braket{x|w_k}\\
&=\sum_{k=1}^{\dim(S^\perp)+1}\sum_{x\in B_0}\order{\braket{w_k|x}}^2\\
&=\sum_{x\in B_0}\order{x}^2\\
&=\dim(S\cap\operatorname{span}(\{w_k\}_1^{\dim(S^\perp)+1}))\\
&\ge\dim(S)+\dim(S^\perp)+1-n\\
&=1
\end{align*} by the subadditivity of codimensions. Then $\braket{w_k|P_S|w_k}$ must be at least $1/(\dim(S^\perp)+1)$ for some $1\le k\le\dim(S^\perp)+1$.\end{proof}

\section{Subgraphs/Colors}

The lemmas above are sufficient to motivate a rigorous definition of subgraphs and coloring by steps. The next lemmas show that the construction is effective by considering the consequence of the separated eigenvalues on possible slopes. The first one demonstrates this efficacy on a subspace that is a code, and the second one demonstrates it on the subspace that is the leftover. Let $G$ be a quantum graph, and let $S$ be a subspace of the vertex space $V(G)$. Then the \emph{induced subgraph} $G|_S$ is the quantum graph such that $V(G|_S)=S$ and $$E(G|_S)=P_SE(G)P_S.$$ Since the error correction condition is unaffected by taking induced subgraphs, any coloring of a quantum graph can be constructed by iteratively taking codes of the subgraph induced by the remaining space. Since classical subgraphs have a quantum equivalent, there is a question of whether other classical graph theoretical concepts can be translated as well. One such concept is that of graph complements. One possible definition is that it is the quantum graph on the same Hilbert space with an edge space that consists of all matrices not contained in the original graph's edge space. The problem with this definition is that the edge space is no longer a linear subspace and does not contain the identity. With a choice of an inner product on matrices, the edge space can be defined as all matrices orthogonal to the original set and including its span with the identity matrix. One such inner product is the trace product defined as $\braket{A,B}=\operatorname{Tr}(AB)$, but other choices are equally as valid.

\begin{lemma}\label{tropical}If $C$ is a code of $Q_{n,m}$, the dimension of $C$ is at most $\ceiling{(\dim(C^\perp)+1)/m}$.\end{lemma}

\begin{proof}By Lemma \ref{basis}, 
$$\braket{w_k|P_C|w_k}\ge1/(\dim(C^\perp)+1)\ge1/n$$
 for some $k\le\dim(C^\perp)+1$. Then $w_k=w_{i,j}$ for some $i$ and some 
$$j\le\ceiling{(\dim(C^\perp)+1)/m}.$$
 By Lemma \ref{rayleigh}, 
$$\lambda_{\max}(P_CA_iP_C)\ge\lambda_{i,j}\braket{w_{i,j}|P_C|w_{i,j}}\ge\lambda_{i,j}/n>\lambda_{i,j+1}.$$
 Then 
$$\epsilon_C(A_i)=\lambda_{\max}(P_CA_iP_C)>\lambda_{i,j+1}.$$
 By Lemma \ref{isotropic}, the dimension of $C$ is at most equal to the cardinality of 
$$\sigma(A_i)_{\ge\epsilon_C(A_i)},$$
 which is at most equal to the cardinality of $\sigma(A_i)_{>\lambda_{i,j+1}}$, which is $j$. Then the dimension of $C$ is at most 
$$\ceiling{(\dim(C^\perp)+1)/m}.$$
\end{proof}

\begin{lemma}\label{cyclical}If $S$ is a nontrivial subspace of $V(Q_{n,m})$, any coloring of $Q_{n,m}|_S$ contains a code of dimension at most $\ceiling{(\dim(S^\perp)+1)/m}$.\end{lemma}

\begin{proof}By Lemma \ref{basis}, 
$$\braket{w_k|P_S|w_k}\ge1/(\dim(S^\perp)+1)\ge1/n$$
 for some $k\le\dim(S^\perp)+1$. Then $w_k=w_{i,j}$ for some $i$ and some 
$$j\le\ceiling{(\dim(S^\perp)+1)/m}.$$
 By Lemma \ref{rayleigh}, $$\lambda_{\max}(P_SA_iP_S)\ge\lambda_{i,j}\braket{w_{i,j}|P_S|w_{i,j}}\ge\lambda_{i,j}/n>\lambda_{i,j+1}n.$$
 Then the mean of $\sigma(P_SA_iP_S)$ is greater than 
$$\lambda_{\max}(P_SA_iP_S)/\dim(S)>\lambda_{i,j+1}n/\dim(S)\ge\lambda_{i,j+1}.$$
 By Lemma \ref{trace}, the multisets $\sigma(P_SA_iP_S)$ and 
$$\epsilon_K(P_SA_iP_S)$$
 have the same mean for any coloring $K$ of $Q_{n,m}|_S$. Then 
$$\epsilon_C(A_i)=\epsilon_C(P_SA_iP_S)>\lambda_{i,j+1}$$
 for some $C\in K$. By Lemma \ref{isotropic}, the dimension of $C$ is at most equal to the cardinality of 
$$\sigma(A_i)_{\ge\epsilon_C(A_i)},$$
 which is at most equal to the cardinality of $\sigma(A_i)_{>\lambda_{i,j+1}}$, which is $j$. Then any coloring of $Q_{n,m}|_S$ must contain a code of dimension at most
$$\ceiling{(\dim(S^\perp)+1)/m}.$$
\end{proof}

\section{Ceilings/Steplogs}

At this point, some combinatorial work is done to rearrange the results into a more useful form. A function is defined in order to more succinctly express them. The next two propositions are the desired results from the construction, and they are half of the main theorems. Define \emph{slog} recursively by 
$$\slog(0,q)=0$$
 and 
$$\slog(p,q)=\slog(p-\ceiling{p/q},q)+1$$
 for any positive integers $p,q$. $\slog$ is short for step logarithm.  I could have alternatively defined the function $\operatorname{stdslog}$ which more closely matches the notation for standard logarithms: 
$$\slog(p,q)=\operatorname{stdslog}_{q/(q-1)}(p)+1\sim\log_{q/(q-1)}(p),$$
 but this does not seem to be a more useful way to represent the function for my purposes. There is a connection between the step logarithm and the Josephus problem, which is as follows. Suppose there are $p$ people in a circle, and every $q$th person is eliminated in a cyclic fashion until there are no more people in the circle. The problem is to determine the initial position of the person who will be eliminated last. Since the problem involves eliminating $\ceiling{k/q}$ people with each traversal of the circle, where $k$ is the number of people remaining, there is an expression involving an inverse of the step logarithm that solves the problem. For a full analysis, see \cite{OW}.

\begin{proposition}\label{ceiling}Every code of $Q_{n,m}$ has dimension at most $\ceiling{n/(m+1)}$.\end{proposition}

\begin{proof}Let $C$ be a code of $Q_{n,m}$. By Lemma \ref{tropical}, \begin{align*}
&\dim(C)\le\ceiling{(\dim(C^\perp)+1)/m}\\\implies
&\dim(C)\le(\dim(C^\perp)+1+x)/m\\\implies
&\dim(C)(m+1)\le\dim(C)+\dim(C^\perp)+1+x\\\implies
&\dim(C)\le(n+1+x)/(m+1)\\\implies
&\dim(C)\le\ceiling{n/(m+1)}
\end{align*} for some integer $0\le x<m$.\end{proof}

\begin{proposition}\label{step}Every coloring of $Q_{n,m}$ has cardinality at least $\slog(n,m+1)$.\end{proposition}

\begin{proof}Define $s_q$ recursively by 
$$s_q(0)=0$$
 and 
$$s_q(p)=s_q(p-1)+\ceiling{(s_q(p-1)+1)/(q-1)}$$
 for any positive integers $p,q$. If $z\in\{0,1\}$, \begin{align*}
&(s_q(p)+z)-\ceiling{(s_q(p)+z)/q}\\=
&\floor{(s_q(p)+z)(q-1)/q}\\=
&\floor{(s_q(p-1)+\ceiling{(s_q(p-1)+1)/(q-1)}+z)(q-1)/q}\\=
&\floor{(s_q(p-1)+(s_q(p-1)+1+x)/(q-1)+z)(q-1)/q}\\=
&s_q(p-1)+z+\floor{(1+x-z)/q}\\=
&s_q(p-1)+z
\end{align*} for some integer $0\le x<q-1$ and for any $p,q$. Then 
$$\slog(s_q(p),q)=\slog(s_q(p-1),q)+1$$
 and 
$$\slog(s_q(p)+1,q)=\slog(s_q(p-1)+1,q)+1$$
 for any $p,q$. Since $\slog$ is nondecreasing, 
$$\slog(s_q(p),q)=\slog(s_q(p-1)+1,q)=\slog(s_q(p-1),q)+1$$
 for any $p,q$. Then, for any $k$, 
$$\slog(k,q)=\slog(s_q(p),q)$$
 for the least $p$ such that $k\le s_q(p)$. By iteratively applying Lemma \ref{cyclical}, every coloring of $Q_{n,m}$ must have cardinality at least $\slog(n,m+1)$.\end{proof}

\section{Partitions/Constructions}

The other half of the main theorems come from a completely different source. A greedy algorithm is used to obtain a subspace that satisfies most of the conditions of being a code. The remaining condition, that it must have constant slope, is satisfied by further refining the subspace. A \emph{partition} is a grouping of elements into disjoint subsets. Given a set of points, Tverberg's theorem \cite{HT} provides a clever way to partition the points into sets that have intersecting convex hulls. The bound obtained below is slightly different from the one by Knill, Laflamme, and Viola due to a minor correction. The original paper has an off by one error that is a result of including and not including the identity in the edge space in different counts. In fact, the counts should be the same, and the denominator is a perfect square.

\begin{proposition}\label{tverberg}Every quantum graph $G$ has a code of dimension at least $\ceiling{\order{G}/(\size{G}+1)^2}$.\end{proposition}

\begin{proof}Take a basis $\{I,E_i\}\subset E(G)$ and a set of orthonormal vectors $\{v_j\}\subset V(G)$ such that 
$$\braket{v_j|E_i|v_k}=\delta_{j,k}c_{i,j}$$
 for all $i,j,k$ and some coefficients $c_{i,j}$. A greedy selection of such vectors gives a set of cardinality at least $\ceiling{\order{G}/(\size{G}+1)}$. By Tverberg's theorem, the set of points $\{(c_{i,j})_i\}_j$ in $\mathbb{R}^{\size{G}}$ can be partitioned into at least $$\ceiling{\ceiling{\order{G}/(\size{G}+1)}/(\size{G}+1)}=\ceiling{\order{G}/(\size{G}+1)^2}$$ sets $P_k$ of indices such that 
$$\sum_{j\in P_k}d_{j,k}c_{i,j}=\sum_{j\in P_l}d_{j,l}c_{i,j}$$
 for all $i,k,l$ and some nonnegative coefficients $d_{j,k}$. Then 
$$\{\sum_{j\in P_k}\sqrt{d_{j,k}}v_j\}_k$$ 
is a basis of the desired code, since 
$$\braket{\sum_{j\in P_k}\sqrt{d_{j,k}}v_j|E_i|\sum_{j\in P_l}\sqrt{d_{j,l}}v_j}=\delta_{k,l}\sum_{j\in P_k}d_{j,k}c_{i,j}$$
 is independent of $k$ for all $i,k,l$.\end{proof}

\begin{proposition}\label{greedy}Every quantum graph $G$ has a coloring of size at most $\slog(\order{G},(\size{G}+1)^2)$.\end{proposition}

\begin{proof}By Proposition \ref{tverberg}, every quantum graph $G$ has a code of dimension at least equal to $\ceiling{\order{G}/(\size{G}+1)^2}$. Iteratively taking this code from subgraphs of $G$ gives a coloring of size at most equal to $\slog(\order{G},(\size{G}+1)^2)$.\end{proof}

\section{Independence/Chromatic}

The propositions so far are so similar that they can be stated jointly with the use of the appropriate definitions. Rather than concerning specific quantum graphs, the main theorems are statements about whole families of quantum graphs with fixed order and valence. The second theorem, in particular, implies the nonexistence of a quantum Brooks' Theorem. Define $\alpha(G)$ and $\chi(G)$ to be the dimension of the largest code and cardinality of the smallest coloring respectively of a quantum graph $G$. Define $\alpha(n,m)$ and $\chi(n,m)$ to be the minimum of $\alpha(G)$ and maximum of $\chi(G)$ respectively among all quantum graphs $G$ such that $\order{G}=n$ and $\size{G}=m$. The space of quantum graphs of order $n$ and valence $m$ forms a subvariety of the Grassmannian $G(n^2,m+1)$, since the edge space is a $m+1$ dimensional subspace of the $n^2$ dimensional real vector space of Hermitian matrices on a Hilbert space of complex dimension $n$. Since each edge space must contain the identity, it is a subvariety and contains an embedding of 
$$G(n^2-1,m).$$
 Then the quantities $\alpha$ and $\chi$ are step functions on the subvariety, and they are lower semicontinuous and upper semicontinuous respectively. This can be shown by taking a limit of codes or colorings to obtain a code or coloring of the limiting quantum graph.

\begin{theorem}\label{independence}For all $n,m$, $$\ceiling{n/(m+1)}\ge\alpha(n,m)\ge\ceiling{n/(m+1)^2}.$$\end{theorem}

\begin{proof}Follows from Proposition \ref{ceiling} and Proposition \ref{tverberg}.\end{proof}

\begin{theorem}\label{chromatic}For all $n,m$, $$\slog(n,m+1)\le\chi(n,m)\le\slog(n,(m+1)^2).$$\end{theorem}

\begin{proof}Follows from Proposition \ref{step} and Proposition \ref{greedy}.\end{proof}

\section{Corollaries/Conjectures}

$$\underline{1}\text{ }\underline{1}\text{ }\underline{1}\text{ }\underline{1}\text{ }\underline{1}\text{ }\underline{22}\text{ }\underline{22}\text{ }\underline{233}\text{ }\underline{3334}\text{ }\underline{4444}$$ 
$$\underline{1}\text{ }\underline{1}\text{ }\underline{1}\text{ }\underline{1}\text{ }\underline{22}\text{ }\underline{22}\text{ }\underline{333}\text{ }\underline{344}\text{ }\underline{4455}\text{ }\underline{55}$$ 
One curiosity arising from these proofs is that $\slog(p,q)$ can be computed in two different ways, as demonstrated above for $p=20$ and $q=5$. By its definition, $\slog(p,q)$ can be computed by iteratively subtracting $1/q$ of the remainder, rounded up, at each step. This is demonstrated by the first line. Alternatively, as discussed in Proposition \ref{step}, $\slog(p,q)$ can be computed by iteratively adding $1/(q-1)$ of the current sum plus one, rounded up, at each step. This is demonstrated by the second line. A shorthand way of doing these computations is to count $p$ digits by repeating each digit $q$ and $q-1$ times respectively. For the first computation, start underlining from the right side, with the number of digits to be underlined given by the rightmost number in the underline. For the second computation, start underlining from the left side, with the number of digits to be underlined given by the leftmost number in the underline. The number of lines drawn in either case will be $\slog(p,q)$, even though the number of lines of each length in each count may be different.

\begin{corollary}\label{commutative}$\alpha(Q_{n,m})=\ceiling{n/(m+1)}$. For quantum graphs $G$ such that $\size{G}=1$, $\chi(G)=\slog(\order{G},2)=\floor{\log_2(\order{G})}+1$.\end{corollary}

Since the edge space of $Q_{n,m}$ is commutative, mutual eigenvectors will satisfy the first half of Proposition \ref{tverberg}. Then only one factor of $1/(\size{G}+1)$ is necessary, resulting from the Tverberg step, and the bounds match. If $\size{G}=1$, $\size{G|_S}\le1$ for any $S$. Then $G|_S$ is isomorphic to $Q_{\size{G|_S},1}$ or $Q_{\size{G|_S},0}$ for any $S$, so codes of size at least $1/(\size{G}+1)=1/2$ of the remaining space can be iteratively removed for a coloring of size $\floor{\log_2(\order{G})}+1$. Conveniently, $\slog(p,q)$ can be expressed with just $\log$ for $q=2$. For classical graphs, $\size{G}=1$ means that $G$ is just a matching. Trivially, every matching has chromatic number $2$, so it is notable that the chromatic number is not even bounded in the quantum case when $\size{G}=1$. Another curiosity is that all results above apply also to infinite quantum graphs, since the construction can easily be extended. In particular, there are infinite quantum graphs of any valence with no finite chromatic number.

\begin{conjecture}\label{constant}For some increasing sequence $1\le c_m<2$ converging to $2$, $\alpha(n,m)=\ceiling{n/(m+1)^{c_m}}$ and $\chi(n,m)=\slog(n,(m+1)^{c_m})$.\end{conjecture}

From the corollary above, $c_m=1$ when $m=1$. The bounds I obtained were a result of spreading out tropicality, and it may be the case that noncommutativity allows for a more pronounced effect. Since the identity always commutes, there is some degree of commutativity for any $m$. If more dimensions dilute this commutativity, the function should be increasing. My complete guess is that $c_m=1$ is totally commutative and $c_m=2$ is totally noncommutative, although that can never be achieved with the required inclusion of the identity, so the sequence $c_m$ tends to $2$ as $m$ approaches infinity. The case of $m=2$ should be telling, and solving that case could go a long way toward proving or disproving this conjecture.


\begin{thebibliography}{00}
\bibitem{RB}R. L. Brooks. ``On colouring the nodes of a network.'' \emph{Mathematical Proceedings of the Cambridge Philosophical Society}. Vol. 37. No. 02. Cambridge University Press, 1941.
\bibitem{CRSS}R. Calderbank, E. M. Rains, P. W. Shor, N. J. A. Sloane. ``Quantum error correction via codes over GF (4).'' arXiv:quant-ph/9608006 (1996).
\bibitem{MC}M. D. Choi. ``Completely positive linear maps on complex matrices.'' \emph{Linear algebra and its applications} 10.3 (1975): 285-290.
\bibitem{DSW}R. Duan, S. Severini, A. Winter. ``Zero-error communication via quantum channels, noncommutative graphs, and a quantum Lov\'{a}sz number.'' \emph{Information Theory, IEEE Transactions on} 59.2 (2013): 1164-1174. arXiv:1002.2514 (2010).
\bibitem{PE}P. Erd\H{o}s. ``Graph theory and probability.'' \emph{Canad. J. Math} 11 (1959): 34G38.
\bibitem{GN}I. M. Gelfand, M. A. Naimark. ``On the imbedding of normed rings into the ring of operators in Hilbert space.'' \emph{Matematiceskij sbornik} 54.2 (1943): 197-217.
\bibitem{RH}R. W. Hamming. ``Error detecting and error correcting codes.'' \emph{Bell System technical journal} 29.2 (1950): 147-160.
\bibitem{KL}E. Knill, R. Laflamme. ``Theory of quantum error-correcting codes.'' \emph{Physical Review A} 55.2 (1997): 900. quant-ph/9604034 (1996).
\bibitem{KLV}E. Knill, R. Laflamme, L. Viola. ``Theory of quantum error correction for general noise.'' \emph{Physical Review Letters} 84.11 (2000): 2525. arXiv:quant-ph/9908066 (1999).
\bibitem{KW}G. J. Kuperberg, N. Weaver. ``A von Neumann algebra approach to quantum metrics.'' No. 1010. \emph{American Mathematical Soc.}, 2012. arXiv:1005.0353 (2010).
\bibitem{LL}L. Lov\'{a}sz. ``On the Shannon capacity of a graph.'' \emph{Information Theory, IEEE Transactions on} 25.1 (1979): 1-7.
\bibitem{MN}F. J. Murray, J. V. Neumann. ``On rings of operators.'' \emph{Annals of Mathematics} (1936): 116-229.
\bibitem{OW}A. M. Odlyzko, H. S. Wilf. ``Functional iteration and the Josephus problem.'' \emph{Glasgow Mathematical Journal} 33.02 (1991): 235-240.
\bibitem{SS}S. Sakai. ``A characterization of W$^*$-algebras.'' \emph{Pacific Journal of Mathematics} 6.4 (1956): 763-773.
\bibitem{CS}C. E. Shannon. ``A mathematical theory of communication.'' \emph{ACM SIGMOBILE Mobile Computing and Communications Review} 5.1 (2001): 3-55.
\bibitem{PS}P. W. Shor. ``Scheme for reducing decoherence in quantum computer memory.'' \emph{Physical review A} 52.4 (1995): R2493.
\bibitem{PT}P. Tur\'{a}n. ``On an extremal problem in graph theory.'' \emph{Mat. Fiz. Lapok} 48.436-452 (1941): 137.
\bibitem{HT}H. A. Tverberg. ``A generalization of Radon's theorem.'' \emph{J. London Math. Soc} 41.1 (1966): 123-128.
\bibitem{NW}N. Weaver. ``Quantum relations.'' arXiv:1005.0354 (2010).
\bibitem{JW}J. H. M. Wedderburn. ``On hypercomplex numbers.'' \emph{Proceedings of the London Mathematical Society} 2.1 (1908): 77-118.
\end{thebibliography}
\end{document}